# Time-efficient, High Resolution 3T Whole Brain Quantitative Relaxometry using 3D-QALAS with Wave-CAIPI Readouts

Jaejin Cho[1,2,†], Borjan Gagoski[2,3,†,*], Tae Hyung Kim[4], Fuyixue Wang[1,2], Daniel Nico Splitthoff[5], Wei-Ching Lo[6], Wei Liu[7], Daniel Polak[5], Stephen Cauley[1,2], Kawin Setsompop[8,9], P. Ellen Grant[2,3,‡], Berkin Bilgic[1,2,10,‡]

[1] Athinoula A. Martinos Center for Biomedical Imaging, Massachusetts General Hospital, Charlestown, MA, United States

[2] Department of Radiology, Harvard Medical School, Boston, MA, United States

[3] Fetal-Neonatal Neuroimaging & Developmental Science Center, Boston Children's Hospital, Boston, MA, United States

[4] Department of Computer Engineering, Hongik University, Seoul, South Korea

[5] Siemens Healthcare GmbH, Erlangen, Germany

[6] Siemens Medical Solutions USA, Inc., Charlestown, Massachusetts, USA

[7] Siemens Shenzhen Magnetic Resonance Ltd, Shenzhen, China

[8] Department of Electrical Engineering, Stanford University, Stanford, California, USA

[9] Department of Radiology, Stanford University, Stanford, California, USA

[10] Harvard/MIT Health Sciences and Technology, Cambridge, MA, United States

[†] Both authors contributed equally to this work as the first authors

[‡] Both authors contributed equally to this work as the last authors

[*] Corresponding author: Borjan Gagoski (borjan.gagoski@childrens.harvard.edu)

**Short title:** Fast whole-brain quantitative imaging using wave-CAIPI 3D QALAS

**ABSTRACT**

**Purpose:** Volumetric, high-resolution, quantitative mapping of brain tissue relaxation properties is hindered by long acquisition times and signal-to-noise (SNR) challenges. This study, for the first time, combines the time-efficient wave-CAIPI readouts into the 3D-quantification using an interleaved Look-Locker acquisition sequence with a $T_2$ preparation pulse (3D-QALAS) acquisition scheme, enabling full brain quantitative $T_1$, $T_2$ and proton density (PD) maps at 1.15 mm$^3$ isotropic voxels in only 3 minutes.

**Methods:** Wave-CAIPI readouts were embedded in the standard 3D-QALAS encoding scheme, enabling full brain quantitative parameter maps ($T_1$, $T_2$, and PD) at acceleration factors of R=3x2 with minimum SNR loss due to g-factor penalties. The quantitative parameter maps were estimated using a dictionary-based mapping algorithm incorporating inversion efficiency and $B_1$ field inhomogeneity. The quantitative maps using the accelerated protocol were quantitatively compared against those obtained from conventional 3D-QALAS sequence using GRAPPA acceleration of R=2 in the ISMRM NIST phantom, and ten healthy volunteers.

**Results:** When tested in both the ISMRM/NIST phantom and ten healthy volunteers, the quantitative maps using the accelerated protocol showed excellent agreement against those obtained from conventional 3D-QALAS at $R_{GRAPPA}$=2.

**Conclusion:** 3D-QALAS enhanced with wave-CAIPI readouts enables time-efficient, full brain quantitative $T_1$, $T_2$, and PD mapping at 1.15 mm$^3$ in 3 minutes at R=3x2 acceleration. When tested on the NIST phantom and ten healthy volunteers, the quantitative maps obtained from the accelerated wave-CAIPI 3D-QALAS protocol showed very similar values to those obtained from the standard 3D-QALAS (R=2) protocol, alluding to the robustness and reliability of the proposed methods.

# 1. INTRODUCTION

Magnetic resonance imaging (MRI) is a powerful tool capable of probing and visualizing the human body non-invasively. Its use in clinical settings is mainly dominated by MR acquisitions that provide a qualitative assessment of tissues' properties, where contrast-weighted images are used to make decisions related to the presence or absence of particular abnormalities. Specifically, neuroimaging protocols routinely include structural acquisitions such as $T_1$-weighted ($T_1$w) MPRAGE (1), $T_2$-weighted ($T_2$w) turbo-spin echo, $T_2$-weighted fluid-attenuated inversion recovery ($T_2$w-FLAIR) (2), or a $T_2^*$-weighted ($T_2^*$w) 3D gradient recalled echo (GRE) from which susceptibility weighted imaging (SWI) (3) can be derived. Furthermore, typical runs of these standard sequences underutilize parallel imaging acceleration available on today's modern hardware, leading to acquisition inefficiency and making structural imaging consume a substantial portion of the scan time budget. Although this mode of operation has been used in clinical settings for decades now, it often prohibits the detection of subtle tissue changes in various pathologies, as the limited set of different contrast-weighted volumes might not provide the contrast needed to differentiate between healthy and abnormal tissue.

To that end, during the past decade, there has been a significant push toward the development of <u>quantitative</u> MRI (qMRI) techniques that aim to provide quantitative estimates of tissue relaxation parameters. These methods (4–9) have developed different pulse sequences with specific, carefully chosen set/range of imaging parameters (e.g. inversion times, echo times, repetition times, etc), and use Bloch simulations of the acquired signals to calculate $T_1$, $T_2$, $T_2^*$ or proton density (PD) maps in quantitative units (percentage for the PD maps, and seconds for the others). In clinical settings, qMRI can help identify physiological changes undetected by qualitative imaging (10), provide specific information to characterize pathologies (11,12), help assess treatment response and repair processes (13), and detect disease before morphological changes (14). qMRI has been applied in epilepsy (15), with some studies showing the correlation between $T_1$ value changes across cortical layers and myelin histological staining in those same brain regions (16,17), while other studies demonstrating that a $T_2$ value increase is concordant with the putative seizure onset (18,19). Similarly, qMRI has been applied in patients with multiple sclerosis, linking disease activity with $T_2$ value changes in normal-appearing white matter (20). Similarly, other studies (21,22) have also shown that $T_1$ relaxation values change in both lesions and white matter in patients with multiple sclerosis and that these changes correlate with increased patients' disabilities (22).

Moreover, qMRI is also beneficial in research settings, as standard qualitative sequences produce system-dependent pixel intensities that cannot be meaningfully compared across sites. Therefore, qMRI is well-suited for multi-center studies, not only because it obtains objective measures of tissue-specific parameters that do not (ideally) depend on the particular sequence or hardware used to obtain them, but also because it has demonstrated significantly higher inter-site reproducibility compared to contrast-weighted imaging (23). In addition, qMRI also provides an elegant solution whereby contrast-weighted images can be derived from calculated relaxation parameter maps (24,25).

Besides these benefits, routine deployment of qMRI in clinical and research studies is still quite limited, mainly because all the current qMRI methods are encoding intensive, suffering from long scan times and/or lower spatial resolution. One of the most time-efficient qMRI methods is the 3D quantification using an interleaved Look-Locker acquisition sequence with $T_2$ preparation pulse (QALAS) (9) - a 3D acquisition employing spoiled turbo-flash readouts with interleaved $T_2$-preparation and inversion recovery RF pulses. Specifically, a typical 3D QALAS acquisition includes five turbo-flash readouts separated 900ms apart, with a 100ms long $T_2$-preparation module preceding the first readout and the remaining four readouts following an inversion pulse that captures $T_1$ dynamics. The 3D-QALAS sequence has shown a strong correlation with reference $T_1$, $T_2$, and PD values in the NIST/ISMRM phantom and high reproducibility in brain imaging (25). The estimated parameter maps further lend themselves to the generation of contrast-weighted images (e.g. $T_1$w, $T_2$w, $T_2$-FLAIR, etc) via Bloch simulation. It has been shown that cortical thickness measurements from synthetically generated (simulated) $T_1$w volumes, obtained from a 12 minutes long QALAS acquisition at 1mm$^3$ isotropic voxels, have high reproducibility and are in agreement with those obtained from a standard MPRAGE (26). Even though the richness of the data obtained in a single 3D QALAS scan is indisputable - $T_1$, $T_2$, PD maps, as well as synthesized volumes that can be easily tuned to different contrast weightings - the overall scan times of over 10 minutes still remain the main hurdle that prevents the widespread use of this technique in both clinical and research studies.

Previous attempts to mitigate the long acquisition times of the 3D QALAS sequence have incorporated compressed sensing acceleration (27), enabling multi-parametric quantitative whole-brain imaging at 1mm$^3$ isotropic voxels in a bit less than 6 minutes. The main aim of this work is to accelerate 3D QALAS even further (while keeping similar voxel sizes), by using an alternative acceleration scheme - *the wave-controlled aliasing in parallel imaging (wave-CAIPI) algorithm* - which has been shown to maximize the signal-to-noise ratios (SNR) by minimizing g-factor ratios (28,29). Specifically, by playing sinusoidal gradients along two spatial axes, the wave-CAIPI encoding scheme spreads the aliasing evenly in all spatial directions, thereby taking full advantage of 3D coil sensitivity distribution, and thus achieving mean and maximum g-factor ratio values of 1.03 and 1.08 at 3T for R = 3×3 acceleration, respectively. Wave-CAIPI has been successfully employed in various different sequences (30–33), including the $T_1$w MPRAGE (34), and it has been successfully deployed in clinical settings, enabling dramatic decreases in the overall scan times compared to the standard, longer clinical protocol, without losing any of the image quality needed for clinical diagnosis (35–38).

This work embeds the time-efficient wave-CAIPI readouts in the 3D QALAS sequence, enabling one to obtain whole brain qMRI in a little over 3 minutes at 1.15mm³ isotropic voxels. It demonstrates that the $T_1$, $T_2$ and PD maps calculated from this accelerated wave-CAIPI QALAS scans show high accuracy when compared with the reference $T_1$, $T_2$ and PD values in the NIST/ISMRM phantom, and are also in good agreement with the equivalent parametric maps obtained from the standard QALAS acquisition (R=2, TA = ~9 minutes) in normal adult volunteers.

## 2. METHODS

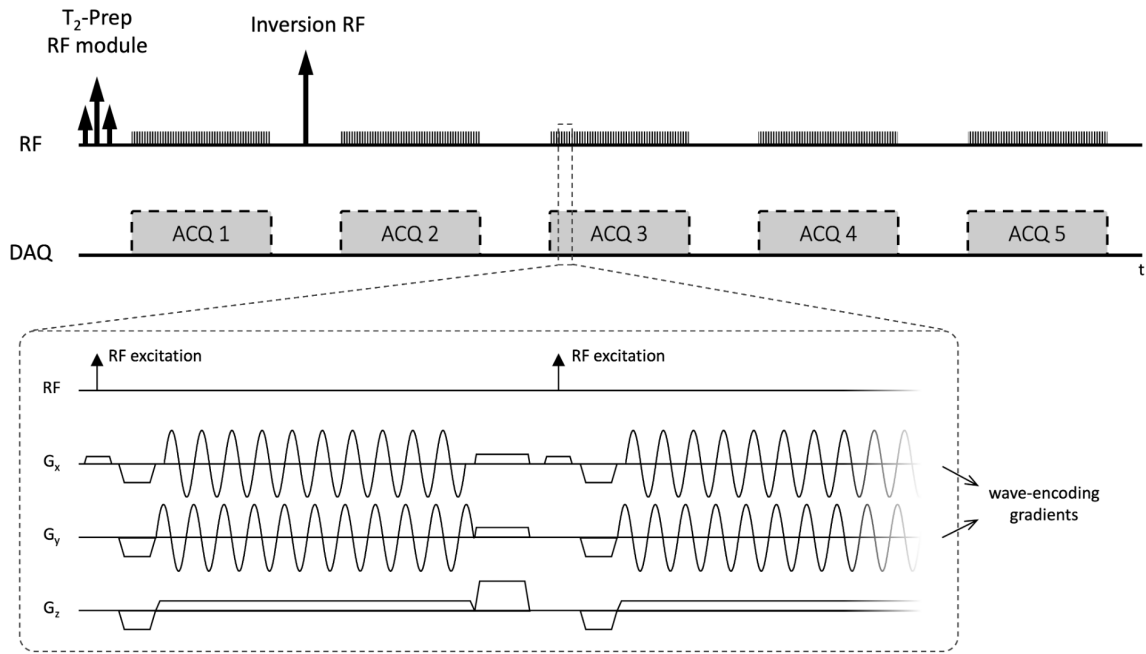

**Figure 1**: Pulse sequence diagram of the proposed wave-CAIPI 3D-QALAS sequence. The top of the figure shows one repetition period, where one can see the $T_2$-prep and inversion RF modules, as well as clearly distinguish the five echo-train readouts, each sampling covering the same portion of k-space in each TR. The bottom of the figure provides a more detailed view of the timing diagram, where one can easily identify the wave-encoded gradient being played along $G_x$ and $G_y$ for a sagittal MRI acquisition.

## 2.1. Wave 3D-QALAS Sequence

The 3D-QALAS acquisition includes five turbo-flash readouts separated 900ms apart, with a 100ms long $T_2$-preparation module preceding the first readout and the remaining four readouts following an inversion pulse that captures $T_1$ dynamics. On the other hand, wave-CAIPI readouts play sinusoidal gradients along two spatial axes, thereby effectively spreading the aliasing evenly in all spatial directions, and thus taking full advantage of 3D coil sensitivity distribution (28,29). Figure 1 shows the sequence diagram of the proposed wave-CAIPI 3D-QALAS sequence. In sagittal imaging, sine and cosine wave gradients are being played in the $G_x$ and $G_y$ directions, respectively.

## 2.2. Experiment

Both phantom and in-vivo experiments were conducted on a 3T MAGNETOM Prisma scanner (Siemens Healthcare, Erlangen, Germany) using a 32-ch head receive array. Wave 3D-QALAS has the following imaging parameters: FOV=240x240x202mm$^3$, matrix size=208x208x176, BW=330 Hz/pixel, echo-spacing=5.9ms, turbo factor=125, TR=4.5s, TE=2.36ms, 10 wave cycles and $R_{yz}$=3x2 acceleration, yielding a scan time of 3:03 minutes. This accelerated protocol was compared against a standard (Cartesian) 3D-QALAS acquisition with equivalent imaging parameters, but with the conventional acceleration of $R_y$=2, resulting in an 8.5 minutes scan. Both 3D-QALAS acquisitions were elliptically acquired. Both acquisitions were run on the ISMRM/NIST phantom, and the two sets of quantitative maps were compared using the Bland-Altman analysis. Similarly, the $T_1$ and $T_2$ maps agreement between the two acquisitions was also evaluated *in vivo* on 10 adult volunteers, by directly comparing the estimated values and performing a Bland-Altman analysis on cortical deep gray and white matter regions. These regions were segmented in FreeSurfer (39) using the synthetically generated $T_1$w volume generated as the following equation.

$$I_{T_1w} = \frac{\rho \sin \alpha \left(1 - exp\left(\frac{T_{TR}}{T_1}\right)\right) exp\left(-\frac{T_{TE}}{T_2}\right)}{1 - \left(\cos \alpha * exp\left(\frac{T_{TR}}{T_1}\right)\right)}$$

where $\rho$, $T_1$, $T_2$ are the estimated PD, $T_1$, and $T_2$ maps, and $\alpha$ is the excitation pulse degree. We used $T_{TE}$=4ms, $T_{TR}$=50ms, and $\alpha$=60° to synthesize T$_1$w images.

## 2.3. Reconstruction and Quantitative Map Estimation

The wave-CAIPI reconstruction used no regularization and was performed online with DICOM images being available on the scanner after the scan was done. For estimating quantitative maps, a dictionary-based matching algorithm was used. In a practical acquisition, RF pulses are not perfectly applied as those are designed. To take into account this, the dictionary was generated for the 50 bins of the inversion efficiency (IE) from 0.75 to 1 and for the 25 bins of the B$_1$ field that was directly acquired from the scanner.

## 3. RESULTS

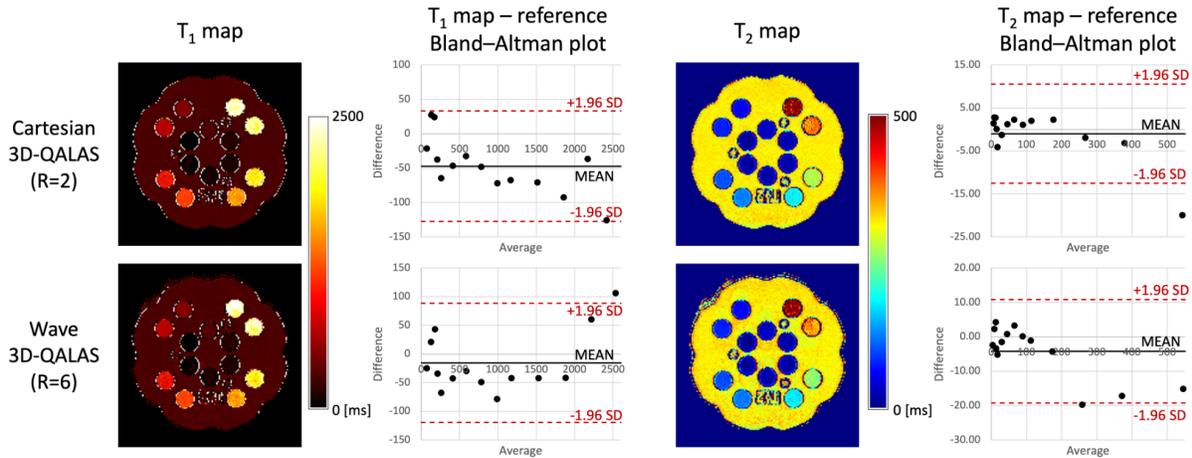

**Figure 2**. Quantitative T$_1$ and T$_2$ maps calculated from standard R=2 and wave-CAIPI R=3x2 3D-QALAS and their Bland-Altman analysis against the reference values. In the Bland-Altman plots, the center black lines represent the mean differences, while the upper and lower red lines represent the 95% confidence interval. Both standard R=2 and wave-CAIPI R=3x2 3D-QALAS results were quantitatively well aligned with the reference values.

Figure 2 shows the $T_1$ and $T_2$ estimates of the ISMRM/NIST phantom from the conventional (R=2) and wave-CAIPI (R=3x2) 3D-QALAS acquisitions. Vials with $T_1$ and $T_2$ values within the brain's physiological range were included in the fit, i.e. between 80 - 2500 ms and 5 - 600 ms for $T_1$ and $T_2$, respectively. Bland-Altman analysis of the estimated $T_1$ and $T_2$ values against the reference values of the phantom shows that the $T_1$ and $T_2$ values are well aligned with the reference values. Most of the difference values are within the 95% confidence interval.

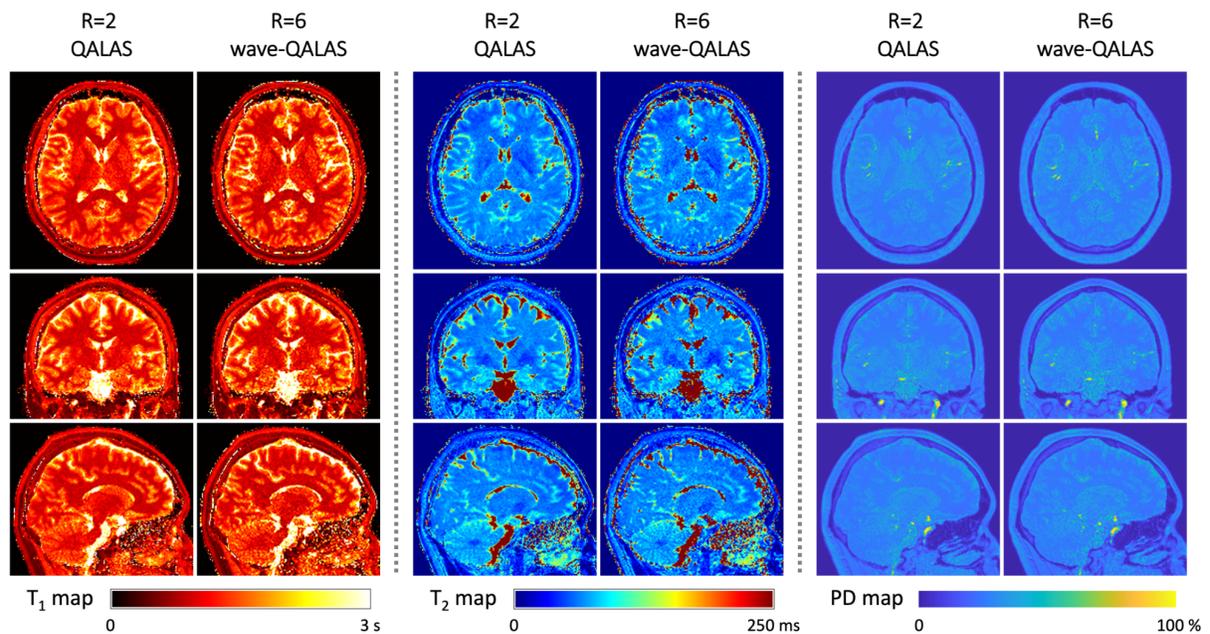

**Figure 3**: Quantitative $T_1$, $T_2$, and PD maps from a representative adult volunteer estimated from the standard Cartesian R=2, and the accelerated R=3x2 wave-CAIPI 3D-QALAS acquisition. Despite the R=3x2 wave-CAIPI scan being almost three times faster, the obtained results show qualitatively similar maps.

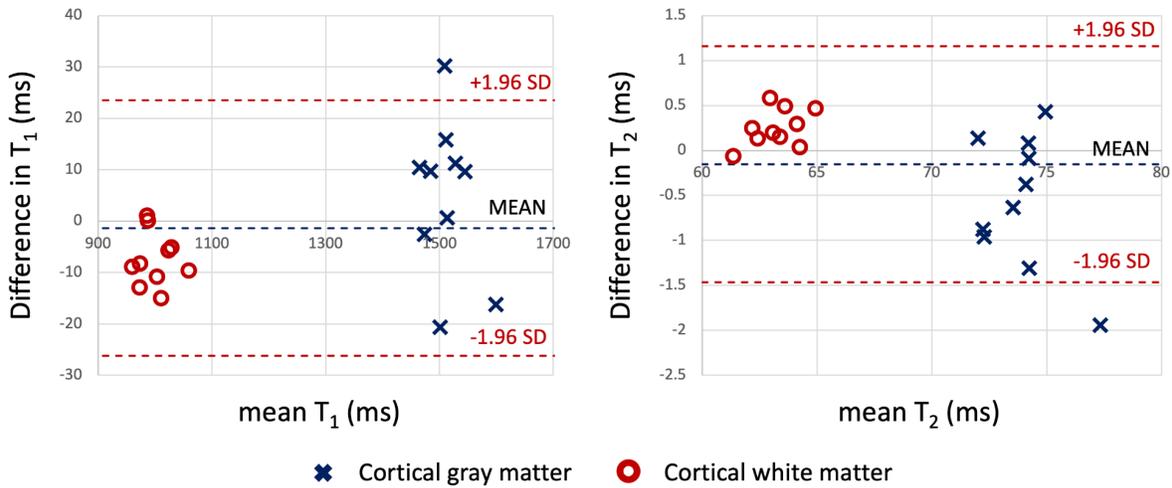

**Figure 4:** Bland-Altman plots comparing estimated $T_1$ and $T_2$ values in the cortical white and deep gray matter in 10 adult volunteers, calculated from standard R=2 and wave-CAIPI R=3x2 3D-QALAS. The center black lines represent the mean differences, while the upper and lower red lines represent the 95% confidence interval. While most points are within the limits of agreement, some minor biases ($T_1$: -1.4ms, $T_2$: -0.15ms) are observed.

Figure 3 shows the $T_1$, $T_2$, and PD maps from both the conventional and wave-CAIPI 3D-QALAS acquisition from a representative subject, demonstrating how similar the two sets of maps are qualitative. Furthermore, Figure 4 shows Bland-Altman analysis of the estimated $T_1$ and $T_2$ values calculated from the two acquisitions in the cortical, white and deep gray matter from all 10 subjects. As seen, the most of $T_1$ and $T_2$ differences between the two acquisitions are within 95% confidence, demonstrating quantitatively that the two acquisitions yield very similar $T_1$ and $T_2$ estimates. Supporting Information Figure S1 shows plot charts of the mean and standard deviations of the $T_1$ and $T_2$ values in these two brain regions in all 10 subjects for both acquisitions.

## 4. DISCUSSION

In this study, we propose to incorporate the wave-CAIPI strategy into the 3D-QALAS acquisition that typically requires a long scan time. Phantom and in-vivo experiments show that wave

3D-QALAS successfully accelerated the scan from 8.5 minutes to 3 minutes at 1.15 mm³ while preserving the accuracy of the quantitative parameter maps.

We used the online DICOM images reconstructed right after the scan without using any regularization. Enabling a quantitative MR exam in less than 5 minutes at 1mm voxels or lower is possible by incorporating advanced reconstruction techniques (40,41), which are part of our future work. Our recent work using neural network denoisers allows a 2-minute exam, but it requires a network training procedure and offline reconstruction.

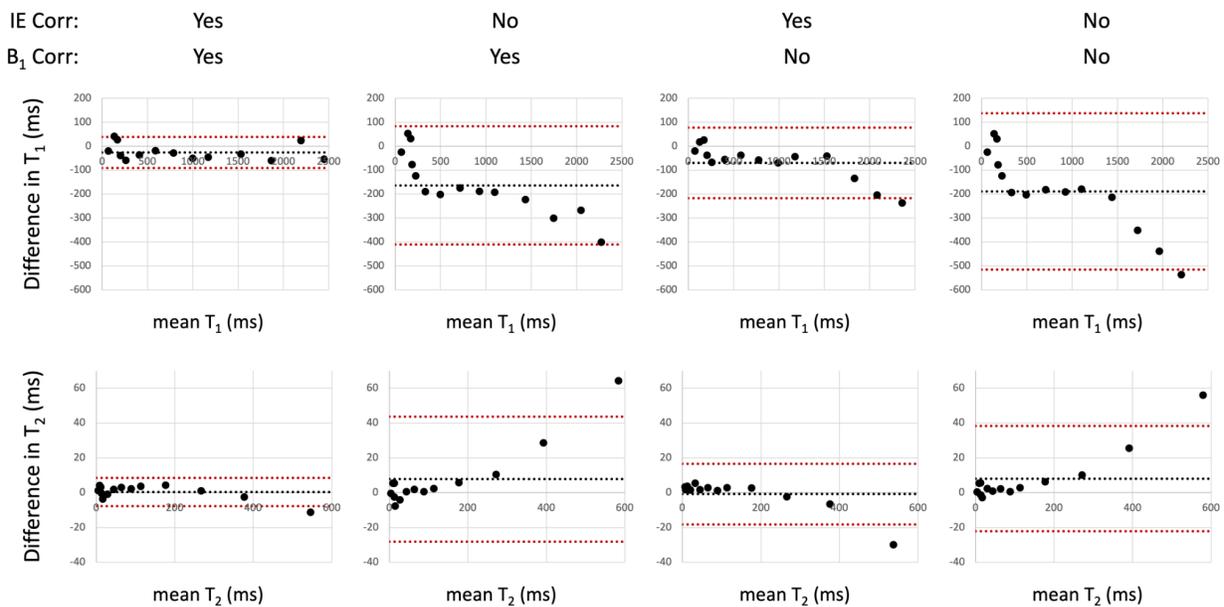

**Figure 5**. Bland-Altman analysis of the estimated $T_1$ and $T_2$ values, calculated from standard R=2 3D-QALAS with/without inversion efficiency (IE) and $B_1$ corrections, against the reference values in the ISMRM-NIST phantom.

To bridge the gap between the theoretical signal model and the practically acquired data, we incorporate IE and the $B_1$ field into the dictionary-based matching algorithm. Figure 5 shows the Bland-Altman analysis of the estimated $T_1$ and $T_2$ values using the standard 3D QALAS (R=2) and the reference values in the ISMRM/NIST phantom with and without IE and the $B_1$ field corrections. IE and $B_1$ field corrections significantly reduced the standard deviation of the difference as well as the mean difference values.

## 5. CONCLUSION

3D-QALAS with wave-CAIPI readouts allows for time-efficient quantitative $T_1$, $T_2$ and PD mapping of the entire brain at 1.15mm$^3$ in 3 minutes, using R=3x2 acceleration. In the quantitative maps in the ISMRM/NIST phantom and 10 adult volunteers, the maps from accelerated wave-CAIPI 3D-QALAS showed very similar values to those obtained from the standard (R=2) 3D-QALAS, alluding to the accuracy and robustness of the proposed methods. The inclusion of IE and $B_1$ field corrections in the dictionary-based matching algorithm further improves the accuracy of $T_1$ and $T_2$ values.

**Acknowledgements:** This work was supported by research grants NIH R01 EB028797, R03 EB031175, U01 EB025162, P41 EB030006, U01 EB026996, R01 EB017337, U01 HD087211, R01 HD100009 and the NVidia Corporation for computing support.

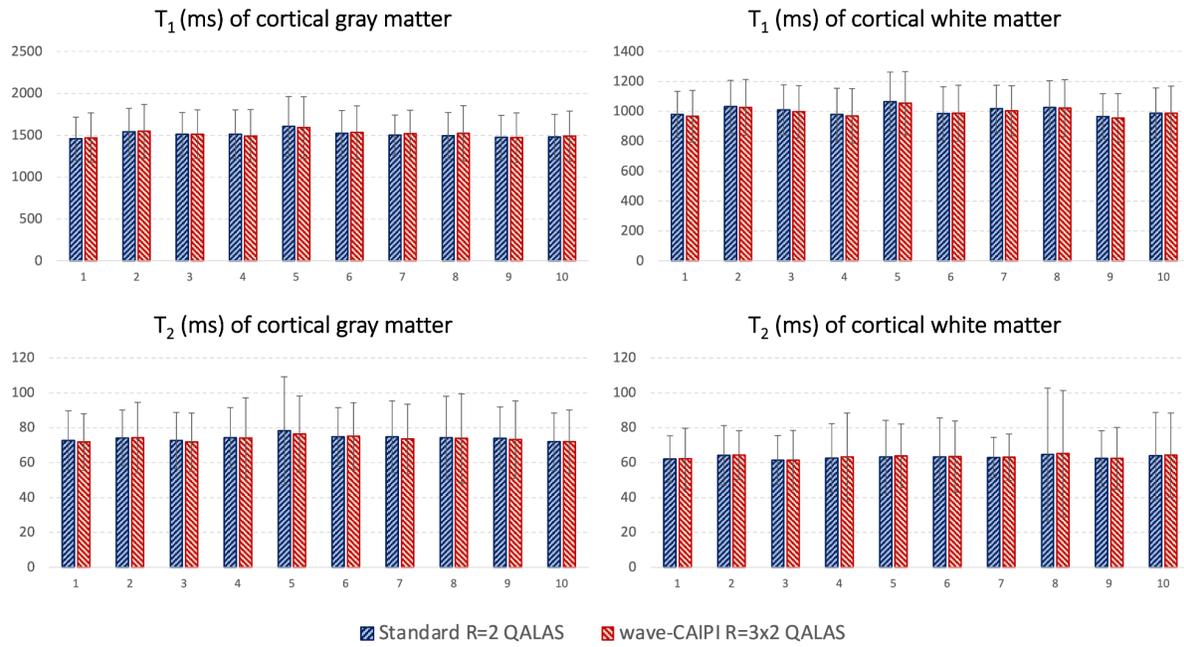

**Supporting Information Figure S1**. Plot charts of the mean and standard deviations of the $T_1$ and $T_2$ values in the cortical white and gray matter regions in all 10 subjects for the standard (R=2) and wave-CAIPI (R=3x2) 3D-QALAS acquisitions.

**REFERENCES**

1. Mugler JP, Brookeman JR. Three-dimensional magnetization-prepared rapid gradient-echo imaging (3D MP RAGE). Magnetic Resonance in Medicine 1990;15:152–157 doi: 10.1002/mrm.1910150117.

2. Mugler JP 3rd. Optimized three-dimensional fast-spin-echo MRI. J. Magn. Reson. Imaging 2014;39:745–767.

3. Haacke EM, Mark Haacke E. Susceptibility Weighted Imaging (SWI). Zeitschrift für Medizinische Physik 2006;16:237 doi: 10.1078/0939-3889-00321.

4. Ehses P, Seiberlich N, Ma D, et al. IR TrueFISP with a golden-ratio-based radial readout: fast quantification of T1, T2, and proton density. Magn. Reson. Med. 2013;69:71–81.

5. Deoni SCL, Peters TM, Rutt BK. High-resolutionT1 andT2 mapping of the brain in a clinically acceptable time with DESPOT1 and DESPOT2. Magnetic Resonance in Medicine 2005;53:237–241 doi: 10.1002/mrm.20314.

6. Fleysher R, Fleysher L, Gonen O. The optimal MR acquisition strategy for exponential decay constants estimation. Magn. Reson. Imaging 2008;26:433–435.

7. Ma D, Gulani V, Seiberlich N, et al. Magnetic resonance fingerprinting. Nature 2013;495:187–192.

8. Liao C, Bilgic B, Manhard MK, et al. 3D MR fingerprinting with accelerated stack-of-spirals and hybrid sliding-window and GRAPPA reconstruction. Neuroimage 2017;162:13–22.

9. Kvernby S, Warntjes MJB, Haraldsson H, Carlhäll C-J, Engvall J, Ebbers T. Simultaneous three-dimensional myocardial T1 and T2 mapping in one breath hold with 3D-QALAS. J. Cardiovasc. Magn. Reson. 2014;16:102.

10. Hamilton-Craig CR, Strudwick MW, Galloway GJ. T1 Mapping for Myocardial Fibrosis by Cardiac Magnetic Resonance Relaxometry—A Comprehensive Technical Review. Frontiers in Cardiovascular Medicine 2017;3 doi: 10.3389/fcvm.2016.00049.

11. MacKay AL, Vavasour IM, Rauscher A, et al. MR Relaxation in Multiple Sclerosis. Neuroimaging Clinics of North America 2009;19:1–26 doi: 10.1016/j.nic.2008.09.007.

12. Liao C, Wang K, Cao X, et al. Detection of Lesions in Mesial Temporal Lobe Epilepsy by Using MR Fingerprinting. Radiology 2018;288:804–812.

13. Welsch GH, Mamisch TC, Domayer SE, et al. Cartilage T2 assessment at 3-T MR imaging: in vivo differentiation of normal hyaline cartilage from reparative tissue after two cartilage repair procedures--initial experience. Radiology 2008;247:154–161.

14. Recht MP, Resnick D. Magnetic Resonance Imaging of Articular Cartilage. Topics in Magnetic Resonance Imaging 1998;9:328???336 doi: 10.1097/00002142-199812000-00002.

15. Ma D, Jones SE, Deshmane A, et al. Development of high-resolution 3D MR fingerprinting for detection and characterization of epileptic lesions. J. Magn. Reson. Imaging 2019;49:1333–1346.

16. Lutti A, Dick F, Sereno MI, Weiskopf N. Using high-resolution quantitative mapping of R1 as an index of cortical myelination. Neuroimage 2014;93 Pt 2:176–188.

17. Waehnert MD, Dinse J, Schäfer A, et al. A subject-specific framework for in vivo myeloarchitectonic analysis using high resolution quantitative MRI. Neuroimage 2016;125:94–107.

18. Rugg-Gunn FJ, Boulby PA, Symms MR, Barker GJ, Duncan JS. Whole-brain T2 mapping demonstrates occult abnormalities in focal epilepsy. Neurology 2005;64:318–325 doi: 10.1212/01.wnl.0000149642.93493.f4.

19. Salmenpera TM, Symms MR, Rugg-Gunn FJ, et al. Evaluation of quantitative magnetic resonance imaging contrasts in MRI-negative refractory focal epilepsy. Epilepsia 2007;48:229–237.

20. Reitz SC, Hof S-M, Fleischer V, et al. Multi-parametric quantitative MRI of normal appearing white matter in multiple sclerosis, and the effect of disease activity on T2. Brain Imaging Behav. 2017;11:744–753.

21. Stevenson VL, Parker GJ, Barker GJ, et al. Variations in T1 and T2 relaxation times of normal appearing white matter and lesions in multiple sclerosis. J. Neurol. Sci. 2000;178:81–87.

22. Parry A, Clare S, Jenkinson M, Smith S, Palace J, Matthews PM. White matter and lesion T1 relaxation times increase in parallel and correlate with disability in multiple sclerosis. J. Neurol. 2002;249:1279–1286.

23. Weiskopf N, Suckling J, Williams G, et al. Quantitative multi-parameter mapping of R1, PD*, MT, and R2* at 3T: a multi-center validation. Frontiers in Neuroscience 2013;7 doi: 10.3389/fnins.2013.00095.

24. Tanenbaum LN, Tsiouris AJ, Johnson AN, et al. Synthetic MRI for Clinical Neuroimaging: Results of the Magnetic Resonance Image Compilation (MAGiC) Prospective, Multicenter, Multireader Trial. American Journal of Neuroradiology 2017;38:1103–1110 doi: 10.3174/ajnr.a5227.

25. Fujita S, Hagiwara A, Hori M, et al. Three-dimensional high-resolution simultaneous quantitative mapping of the whole brain with 3D-QALAS: An accuracy and repeatability study. Magn. Reson. Imaging 2019;63:235–243.

26. Fujita S, Hagiwara A, Hori M, et al. 3D quantitative synthetic MRI-derived cortical thickness and subcortical brain volumes: Scan-rescan repeatability and comparison with conventional T-weighted images. J. Magn. Reson. Imaging 2019;50:1834–1842.

27. Fujita S, Hagiwara A, Takei N, et al. Accelerated Isotropic Multiparametric Imaging by High Spatial Resolution 3D-QALAS With Compressed Sensing: A Phantom, Volunteer, and Patient Study. Invest. Radiol. 2021;56:292–300.

28. Bilgic B, Gagoski BA, Cauley SF, et al. Wave-CAIPI for highly accelerated 3D imaging. Magnetic Resonance in Medicine 2015;73:2152–2162 doi: 10.1002/mrm.25347.

29. Cauley SF, Setsompop K, Bilgic B, Bhat H, Gagoski B, Wald LL. Autocalibrated wave-CAIPI reconstruction; Joint optimization of k-space trajectory and parallel imaging reconstruction.

Magn. Reson. Med. 2017;78:1093–1099.

30. Polak D, Cauley S, Huang SY, et al. Highly-accelerated volumetric brain examination using optimized wave-CAIPI encoding. J. Magn. Reson. Imaging 2019;50:961–974.

31. Bilgic B, Xie L, Dibb R, et al. Rapid multi-orientation quantitative susceptibility mapping. Neuroimage 2016;125:1131–1141.

32. Gagoski BA, Bilgic B, Eichner C, et al. RARE/turbo spin echo imaging with Simultaneous Multislice Wave-CAIPI. Magn. Reson. Med. 2015;73:929–938.

33. Cho J, Liao C, Tian Q, et al. Highly accelerated EPI with wave encoding and multi-shot simultaneous multislice imaging. Magn. Reson. Med. 2022;88:1180–1197.

34. Polak D, Setsompop K, Cauley SF, et al. Wave-CAIPI for highly accelerated MP-RAGE imaging. Magn. Reson. Med. 2018;79:401–406.

35. Goncalves Filho ALM, Longo MGF, Conklin J, et al. MRI Highly Accelerated Wave-CAIPI T1-SPACE versus Standard T1-SPACE to detect brain gadolinium-enhancing lesions at 3T. J. Neuroimaging 2021 doi: 10.1111/jon.12893.

36. Goncalves Filho ALM, Conklin J, Longo MGF, et al. Accelerated Post-contrast Wave-CAIPI T1 SPACE Achieves Equivalent Diagnostic Performance Compared With Standard T1 SPACE for the Detection of Brain Metastases in Clinical 3T MRI. Front. Neurol. 2020;11:587327.

37. Longo MGF, Conklin J, Cauley SF, et al. Evaluation of Ultrafast Wave-CAIPI MPRAGE for Visual Grading and Automated Measurement of Brain Tissue Volume. AJNR Am. J. Neuroradiol. 2020;41:1388–1396.

38. Conklin J, Longo MGF, Cauley SF, et al. Validation of Highly Accelerated Wave-CAIPI SWI Compared with Conventional SWI and T2*-Weighted Gradient Recalled-Echo for Routine Clinical Brain MRI at 3T. AJNR Am. J. Neuroradiol. 2019;40:2073–2080.

39. Fischl B. FreeSurfer. Neuroimage 2012;62:774–781.

40. Bilgic B, Kim TH, Liao C, et al. Improving parallel imaging by jointly reconstructing multi-contrast data. Magn. Reson. Med. 2018;80:619–632.

41. Aggarwal HK, Mani MP, Jacob M. MULTI-SHOT SENSITIVITY-ENCODED DIFFUSION MRI USING MODEL-BASED DEEP LEARNING (MODL-MUSSELS). Proc. IEEE Int. Symp. Biomed. Imaging 2019;2019:1541–1544.